\documentclass[webconf, onecolumn]{svjour}
\usepackage{graphicx}
\usepackage[varg]{txfonts} 
\usepackage[latin1]{inputenc}
\session-title{AO4ELT}
\begin{document}
\title{The physics of galaxy evolution with EAGLE} \author{M.
  Puech\inst{1}\fnmsep\thanks{\email{mathieu.puech@obspm.fr}} \and M.
  Lehnert\inst{1} \and Y. Yang\inst{1} \and J.-G. Cuby\inst{2} \and S.
  Morris\inst{3} \and C. Evans\inst{4} \and B. Neichel\inst{5,1,6}
  \and T. Fusco\inst{5} \and G. Rousset\inst{7} \and H.
  Flores\inst{1}}
\institute{GEPI, Observatoire de Paris, 5 Place Jules Janssen, 92195
  Meudon Cedex, France \and LAM, OAMP, 38 rue Frederic Joliot
  Curie, 13388 Marseille Cedex 13, Francee \and Department of Physics,
  Durham University, South Road, Durham, DH1 3LE, UK \and UK ATC,
  Royal Observatory Edinburgh, Blackford Hill, Edinburgh, EH9 3HJ, UK
  \and ONERA, BP 72, 92322 Chatillon Cedex, France \and Gemini
  Observatory, Colina El Pino s/n, Castilla 603, La Serena, Chile \and
  LESIA, Observatoire de Paris, 5 Place Jules Janssen, 92195 Meudon
  Cedex, France}
\abstract{One of the prominent science goal of the ELTs will be to
  study the physics and mass assembly of galaxies at very high
  redshifts. Here, we present the galaxy evolution science case for
  EAGLE, which is a NIR multi-integral field spectrograph
  for the E-ELT currently under phase A study. We summarize results
  of simulations conducted to derive high-level requirements. In
  particular, we show how we have derived the specifications for the
  ensquared energy that the AO system needs to provide to reach the
  scientific goals of the instrument. Finally, we present future
  strategies to conduct galaxy surveys with EAGLE.}
\maketitle
\section{Introduction}
\label{intro}
EAGLE is a multi-IFU, near-IR spectrograph assisted by Multi-Object
Adaptive Optics (MOAO) for the E-ELT. It is a French-UK partnership
currently under a phase A (design study) in collaboration with ESO.
The current conceptual design is presented in detail elsewhere in
these proceedings \cite{evans,rousset,schnetler}. Briefly, EAGLE has a
patrol field of with an equivalent diameter of 7-arcmin, within which
20, 1.65$\times$1.65 arcsec$^2$ FoV IFUs can be deployed. Image
slicers will provide 37.5 mas slices. EAGLE will cover from 0.8 to 2.5
$\mu$m at a spectral resolution of 4000 in its low resolution, or
10000 in its high resolution mode. A MOAO (Multi-Object Adaptive
Optics) system will deliver excellent image quality using an array of
up to 6 LGS and 5 NGS, correcting for atmospheric turbulence using
deformable mirrors within the instrument and the large deformable
mirror in the telescope itself. An on-sky demonstrator, called CANARY,
is under development in parallel to the EAGLE study
\cite{morris,vidal}.

The conceptual design is driven by five  top-level science cases:
\begin{itemize}
\item Physics and evolution of high-redshift galaxies
\item Detection and characterisation of `first light' galaxies
\item Galaxy assembly and evolution from stellar archaeology
\item Star-formation, stellar clusters and the initial mass function
\item Co-ordinated growth of black holes and galaxies
\end{itemize}

Here, we focus on the first science case. A summary of EAGLE
capabilities relative to the second science case is given by Evans et
al. \cite{evans}, while Paumard et al. illustrate the E-ELT
capabilities relative to the two last cases \cite{paumard}.

\section{Understanding galaxy evolution with EAGLE}
One of the main challanges in extragalactic astronomy is to understand
how galaxies formed and evolved. One of the main issues is to trace
mass assembly in galaxies as a function of cosmic time. Different
physical mechanisms are known to drive this evolution. Major mergers
are collisions between galaxies of similar mass, which can strongly
enhance the conversion of gas into stars in the two progenitors.
Another channel of driving star formation is through minor mergers,
which are galaxy collisions between progenitors with mass ratio of
less than 1:3. Therefore, these minor collisions individually have a
much weaker effect on the mass assembly of a given galaxy, but they
are thought to occur more frequently over its lifetime. Another
channel is direct cold gas accretion from intergalactic filaments. We
do not understand yet what is the dominant channel as a function of
mass and time.

\begin{figure}
\centering
\includegraphics[width=14.3cm]{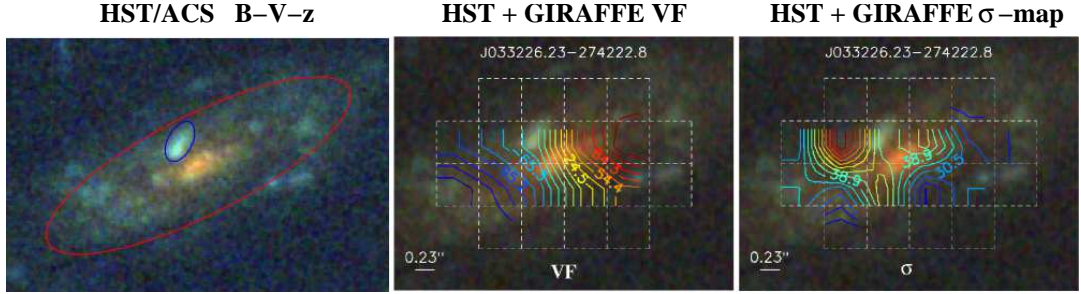}
\caption{3D spectroscopy of a minor merger occuring in a z$\sim$0.6
  star-forming galaxy. \emph{From left to right:} $HST$/ACS $BVz$ color
  image. The red ellipse marks the main progenitor, while the blue
  ellipses indicates the infalling satellite; GIRAFFE velocity fields
  supersimposed in the $HST$ color image; GIRAFFE velocity dispersion
  map superimposed on the $HST$ color image. The white dash-lines in the
  two last panels represent the GIRAFFE IFU pixels.}
\label{giraffe}
\end{figure}

3D spectroscopy can now routinely derive the spatially-resolved
kinematics of massive distant galaxies, up to z$\sim$3. This has
allowed us to make a major breakthrough in our understanding of
distant galaxies, and in particular of what drives star formation in
distant objects. In Fig. \ref{giraffe}, we show one example of how 3D
spectroscopy can help in disentangling various physical processes.
This $HST$/ACS color image (on the left) reveals a blue elongated
region (see the blue ellipse) superimposed on a z$\sim$0.6 galaxy. The
kinematic maps (obtained using GIRAFFE at the VLT, a high spectral
resolution optical spectrograph with multi-IFUs amongst other
capabilities \cite{messenger}) do not show any particular perturbation
in the velocity field of this galaxy, which seems to be a normal
rotating spiral galaxy. On the contrary, the velocity dispersion map
is not what is expected from a regular unperturbed rotating galaxy,
since its peak of dispersion (see the red isocontours) is off-centered
compared to the dynamical centre. This peak falls very close to the
blue elongated region seen on the $HST$ color image. These
observations are exactly what is expected form a minor merger occuring
in a rotating disk: the gas expelled from the infalling satellite
during the merging produces shocks when it encounters the gas
contained in the main progenitor. The off-centered peak of the
velocity dispersion is a kinematic signature of this process.

These observations illustrate how 3D spectroscopy is a quantitative
tool when one wants to disentangle physical processes driving star
formation in distant galaxies. By mapping the physical and chemical
properties of galaxies all the way up to z$\sim$6, EAGLE will the best
tool for studying mass assembly in distant galaxies.

\section{Determining the requirements for the EAGLE IFUs}
The main difficulty in defining the characteristics of an IFU for
EAGLE is that all the physical processes that one wants to study are
operating on very different spatial scales. As shown in Fig.
\ref{giraffe}, minor mergers lead to spatial signatures on the kpc
scale. Major mergers are more violent processes that can destroy
orbital motions, i.e., rotation in disks, which therefore have spatial
signatures on larger scales ($\sim$10-100 kpc). On the other hand,
cold gas accretion can feed galaxies with fresh gas on different
scales, depending on mass and redshift. To elucidate these processes,
one has to adapt the ``scale-coupling'' of the IFU. The scale-coupling
is the ratio between the spatial scale to be resolved by the IFU
(i.e., galaxy diameter for major mergers), to the size of the spatial
element of resolution. In most cases, MOAO will deliver a PSF with
a FWHM smaller than two IFU spaxels. This means that the element of
spatial resolution is almost always driven by the spaxel size (i.e.,
two spaxels), and not directly by the PSF \cite{puech08}.

The scale-coupling directly drives the precision on the physical
quantity one wants to study. For instance, a modest scale-coupling is
already enough to recover the rotation velocity of distant galaxies,
while a much finer spatial resolution is needed if one wants to
recover the whole shape of the rotation curve \cite{epinat09}, which
is needed to study mass profiles. Moreover, once one has decided what
is the optimal scale-coupling, one has to take care of the cross-talk
induced by the PSF wings. Indeed, AO corrections leave a residual halo
sourrounding a diffraction-limited core, which results in a mixing of
the spectral information coming from adjacent spectra on the detector.
Therefore, it is also important to characterize what level of PSF
contrast one needs to recover a specific physical quantity, which
directly implies how easy and accurate it is to distinguish discrete
features over the spatial sampling \cite{puech08}. Both the
scale-coupling and the contrast can be parameterised using the
Ensquared Energy (EE), which is defined as the energy of the PSF
entering one element of spatial resolution.

\begin{figure}
\centering
\includegraphics[width=14.5cm]{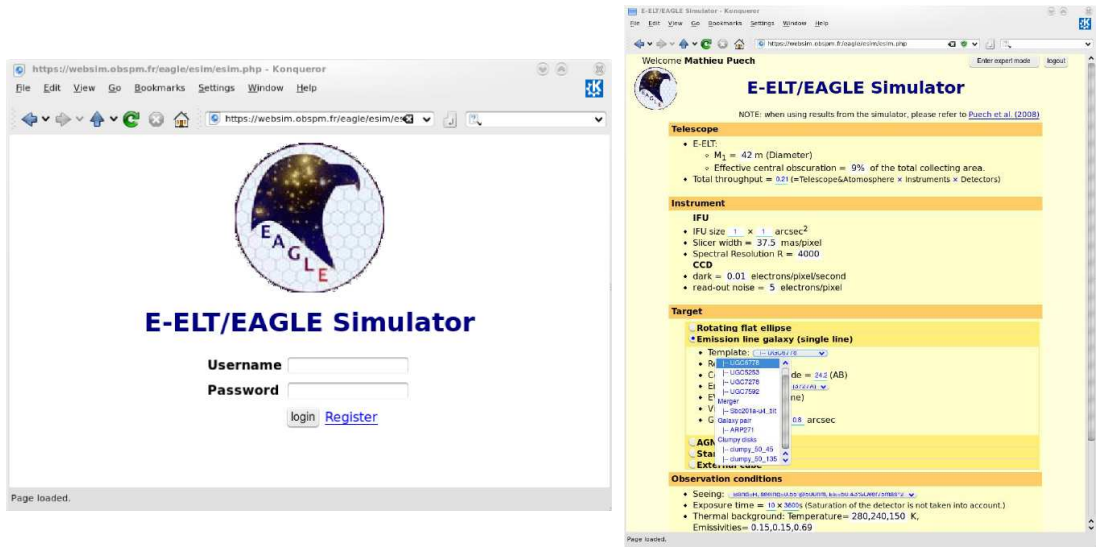}
\caption{Web pages of the EAGLE simulator. The user can pick up a
  science case using a menu (sed right-hand panel) and modify the
  parameters corrsponding to the telescope, instrument, object, sky,
  and detector, as listed.}
\label{websim}
\end{figure}

Determining the optimal scale and EE for the EAGLE IFU is not a
straighforward task. We have therefore developed a tool which allows
us to simulate observations of distant galaxies in a very realistic
way. This simulation tool, with web interface, allows the EAGLE
science team to log-on and run simulations to constrain the
requirements of their specific science case (see Fig. \ref{websim}).
Details about the simulations pipeline can be found elsewhere
\cite{puech08,puech08b}. This is a modified version of the pipeline
used to conduct the design reference mission of the E-ELT for the
galaxy evolution science case \cite{puech09}. This pipeline produces
simulated EAGLE datacubes that can then be analyzed using the same
tools as for real data. From these scientific products (e.g., a
velocity field), one can assess what is, e.g., the required sampling,
or the transmission required to reach a given scientific goal.

\section{What is the relevant spatial scale for studying galaxy evolution?}
\begin{figure}
\centering
\includegraphics[width=9cm]{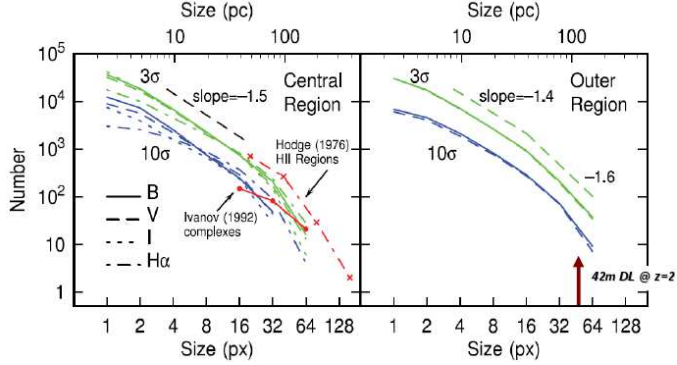}
\caption{Histograms of HII regions in the local galaxy NGC 628, both
  in the central and outer regions \cite{elmegreen06}. The red arrow
  in the bottom of the right panel indicates the limit of diffraction
  of the E-ELT at z=2.}
\label{elm}
\end{figure}

In Fig. \ref{elm}, we show a histogram of the size of HII regions in
local galaxies from Elmegreen et al. \cite{elmegreen06}. 
The E-ELT will not be able to resolve individual HII regions
in very distant objects (z$\sim$2 and farther), with only the largest HII
complexes resolved.  What then is the
optimal spatial scale for studying star formation in distant galaxies?

$HST$ images of cosmological fields have revealed that the morphology
of distant galaxies is much more complex than in the local Universe.
In particular, kpc-sized clumps appear to be ubiquitous in galaxies at
z$>$1 \cite{elmegreen05}. These clumps have sizes of 120-160 mas over
the redshift range 1-5, which means that an EAGLE slice width of
37.5mas is particularly well-suited to study these clumps, offering a
scale-coupling of 3-4. Such a coupling will allow us to detect
rotation in these clumps, which is crucial to understand whether these
clumps are a result of Jeans-instabilites in distant gas-rich disks
fed by cold streams. Indeed, recent numerical simulations have shown
how this process could be an important channel for the formation of
bulges and disks \cite{bournaud,dekel}.

\section{Simulations of distant galaxies: Specifications for the EE}
\begin{figure}[!h]
\centering
\includegraphics[width=14.5cm]{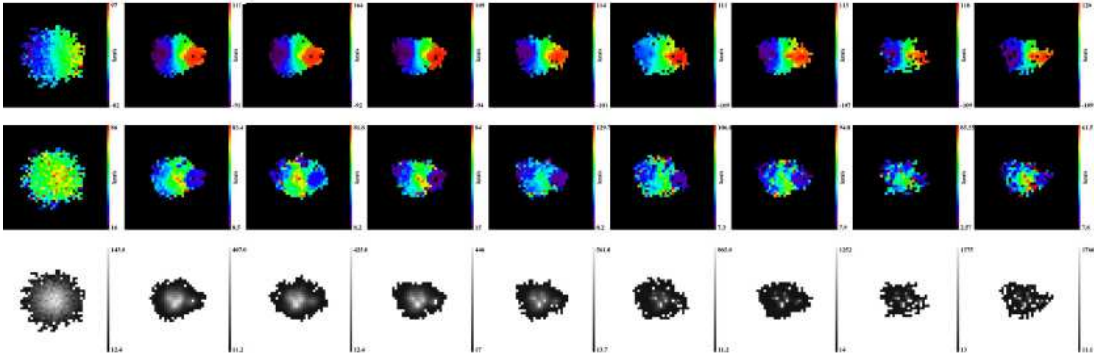}
\caption{Simulations of a clumpy disk at z=4. \emph{From top to
    bottom:} velocity field in the gas (approaching motions are in
  blue, while receding motions are in red), velocity dispersion map,
  and [OII] emission line map. The different columns correspond to
  increasing EE. The EE in 75 mas is 15\% for the second column, while
  it is 61\% in the last one. The first column shows the
  seeing-limited case for comparison. In all simulations, the pixel
  scale is 37.5 mas, and the spectral resolution is 4000. Images are
  0.8$\times$0.8 arcsec$^2$. These simulations were run with a
  conservative Equivalent Width EQW=30\AA. It is therefore a
  challenging experiment to retrieve clumps in these galaxies, since
  currently observed galaxies at z$\sim$2 in such objects have
  EQW$\sim$100\AA.}
\label{clumpy}
\end{figure}

Using the simulation pipeline, we obtained simulated kinematic maps of
a z=4 Jeans-unstable disk, a merging pair of Sbc galaxies, and a
regular rotating disk\footnote{We are especially indebted to T.~J.~Cox
  and F.~Bournaud who provided us with the hydro-dynamical simulations
  of merging galaxies and clumpy disks respectively, and to P.~Amram
  and B.~Epinat who have provided us with kinematical data of local
  galaxies from the GHASP survey.}. The last cases were used to derive
the optimal scale and EE required to distinguish a major merger of a
regular rotation. This case is discussed in detail elsewhere
\cite{puech08}. Here, we focus on the simulations of the
Jeans-unstable clumpy disk, which is shown in Fig. \ref{clumpy}. From
these simulations, one can see that clumps become detectable with at
least $\sim$20\% of EE in 75 mas, while 30\% provides us with a more
robust measurement in case of limited signal-to-noise ratio.

The resulting EE specifications for EAGLE, as a function of a given
scientific goal, are summarized in Tab. \ref{spec1}.

\begin{table}[!h]
\centering
\caption{Constraints on the EE derived from simulations, depending on
  the scientific objective.}
\begin{tabular}{|c|c|c|c|}\hline
Spatial scale & Minimal EE & Optimal EE & Comments\\\hline
Large-scale motions & 30\% in 150 mas  & 30\% in 100 mas  & Good confidence level\\
 & ($\sim$14\% in 75mas) & ($\sim$25\% in 75mas) &\\\hline
Clump detection & 20\% in 75 mas & 30\% in 75 mas & Structure of clumps uncertain\\\hline
\end{tabular}
\label{spec1}
\end{table}

Finally, a trade-off was made between the two science goals, taking account on
the confidence level of all constraints, which is presented in Tab.
\ref{spec2}. This table justifies why the current baseline for EAGLE
assumes 30\% of EE in 75 mas.

\begin{table}[!h]
\centering
\caption{EE Specifications for EAGLE.}
\begin{tabular}{|c|c|c|c|}\hline
 & Minimal & Optimal & Goal\\\hline
EE required in 75 mas & 15\% & 25\% & 30\%\\\hline
\end{tabular}
\label{spec2}
\end{table}

\section{Conducting surveys of high-z galaxies with EAGLE}
Understanding the physical mechanisms driving galaxy evolution will
require 3D spectroscopy of a representative sample of galaxies at
different redshifts. So far, only two such samples have been studied
extensively with 3D spectroscopy: the IMAGES sample, with 100 galaxies
at 0.4$<$z$<$1 using FLAMES/GIRAFFE at the VLT \cite{images}, and the
SINS sample, with 63 galaxies at 1.3$<$z$<$2.6, using VLT/SINFONI
\cite{sins}. Both samples are thought to be approximately
representative of galaxies more massive than $\sim$
10$^{10}$M$_\odot$. Other samples at 2$<$z$<$3 and z$\sim$1.6 were
studied using OSIRIS on the Keck telescope \cite{law,wright}, but they
are drastically limited in size and/or whether they are representative
in their distribution of stellar mass remains unclear.

Such samples clearly allowed us to make an impressive leap forward
into our understanding of galaxy evolution and formation. However, we
are already reaching the limits of what can be done on a 8-10m
telescopes, even equipped by state-of-the-art adaptive optics systems.
This is particularly true at z$>$1, where emission lines are observed
in the NIR H and K bands. In these regions, effective areas of the
spectrum where emission lines can be observed without contending with
strong night-sky lines are limited. In the coming years, the multiplex
advantage of KMOS at VLT will allow us to compile significantly larger
samples of galaxies studies with IFUs at z$>$1. But the 
collecting area of the current generation of telescopes and seeing
limited performance will still limit the size of these samples, since
objects will have to be selected such that their emission lines do not
overlap with strong night-sky lines or strong atmospheric absorption; 
these observations already take several hours on such telescopes. Last
but not least, current distant 3D surveys at z$>$2 contain galaxies
selected in a variety of ways (e.g., BzK, Lyman-Break galaxies, etc.),
which might result in non-trivial and perhaps uncontrollable biases.

EAGLE on the E-ELT will be a decisive step in removing these
limitations. The huge collecting area of the E-ELT and moderate
spectral resolution and high multiplex of EAGLE will allow us to select
galaxies with a less constrained set of redshifts or colour
pre-selection. Potentially, such a combination will allow us to
observe all galaxies (down to a limiting magnitude) at almost any
redshift, which is the only method to obtain a truly representative, robust
sample of galaxies.

In an first example observing programme, EAGLE will enable us to
conduct a broad shallow survey of galaxies, using diagnostic
rest-frame optical emission lines from z$\sim$0.5 to 5, principally
designed to obtain their kinematics, emission line and continuum
morphologies, and clues to the sources of ionization of their ISM.
These observations will take about 8-10 hours per field and will
observe several hundred galaxies. This part of the programme will take
about 100 hours in total. For sources in most favorable redshift
ranges of 1.2-1.7, 2-2.6, and 3-3.6, the rest-frame optical emission
lines are available in three bands (only two for the highest redshift
range listed). In these ranges, we can observe the important
diagnostic lines of [OII], [OIII], [NII], H$\alpha$, and H$\beta$ as
well as other important lines. Such a program will require tens of
hours per field of integration time. For 20 galaxies in each redshift
range will therefore require 240 hours (30 hours per band). Such deep
integrations will enable us to investigate their spatially-resolved
metallicities and ionization mechanisms and provide a detailed
understanding the physical properties of the warm ionized gas in
distant galaxies \cite{lehnert}.


Current spectroscopic surveys already provide us with redshift
catalogues from which EAGLE targets can be drawn. However, these surveys do
not have the necessary spectroscopic completeness 
over several fields of 20$\times$20 arcmin$^2$ in size or greater.
Deeper, wider surveys could be undertaken with VIMOS but we emphasize
the need for good completeness as current surveys generally do not
have the requisite target densities down to faint magnitudes/low
emission equivalent-widths. ALMA may play a role in this by surveying
fields with broad-band receivers in order to obtain redshifts and
dynamics of distant galaxies, especially the dusty, optically faint
galaxies. Also, the multi-slit mode of $JWST$-NIRSPEC could provide
sufficient target densities and high completeness necessary for our
proposed surveys. 

%

\end{document}